\documentclass[hyper,11pt,letterpaper]{JHEP3}

\usepackage{ifpdf}
\usepackage{graphicx}
\usepackage{amsmath, amssymb}
\usepackage{amsfonts}
\usepackage{amsbsy} 

\newcommand{\ba}{\begin{align}}
\newcommand{\ea}{\end{align}}
\newcommand{\bea}{\begin{eqnarray}}
\newcommand{\eea}{\end{eqnarray}}
\newcommand{\be}{\begin{eqnarray}}
\newcommand{\ee}{\end{eqnarray}}
\newcommand{\nn}{\nonumber}
\newcommand{\bn}{\begin{enumerate}}
\newcommand{\en}{\end{enumerate}}

\def\half{\frac{1}{2}}

\def\IC{\mathbb{C}}

\def\IP{\mathbb{P}}
\def\IR{\mathbb{R}}
\def\IZ{\mathbb{Z}}

\def\CM{{\cal M}}
\def\CN{{\cal N}}
\def\CO{{\cal O}}

\def\CS{{\cal S}}

\def\CW{{\cal W}}

\def\a{\alpha}
\def\b{\beta}

\def\e{\epsilon}

\def\Sym{{\rm Sym}}
\def\Alt{{\rm Alt}}

\title{Counting Exceptional Instantons}

\author
{
Christoph A.~Keller\footnote{ckeller@theory.caltech.edu}\ , Jaewon Song\footnote{jaewon@theory.caltech.edu}
\\
\\
California Institute of Technology, Pasadena, CA 91125, USA
}

\abstract
{
We show how to obtain the instanton partition function
of $N=2$ SYM with exceptional gauge group $EFG$ using blow-up recursion relations derived 
by Nakajima and Yoshioka.
We compute the two instanton contribution and match it with the recent proposal 
for the superconformal index of rank 2 SCFTs with $E_{6, 7}$ global symmetry. 
}

\preprint{
CALT-68-2872
}

\begin{document}

\addtolength{\parskip}{.5mm}
\addtolength{\baselineskip}{.2mm}
\addtolength{\abovedisplayskip}{.8mm}
\addtolength{\belowdisplayskip}{.8mm}

\section{Introduction}
For $N=2$ supersymmetric gauge theories, instanton contributions can
be computed exactly using localization. Localization reduces
the infinite dimensional path integral to a finite dimensional
integral over the moduli space of self-dual connections of the
gauge group $G$. 
For classical gauge groups this moduli
space can be described by the ADHM construction \cite{Atiyah:1978ri}.
Applying equivariant localization \cite{Moore:1997dj, Nekrasov:2002qd} to this,
the instanton partition function for  $ABCD$ gauge groups was computed in \cite{Nekrasov:2004vw, Marino:2004cn, Fucito:2004gi}.

From a string theoretic point of view the ADHM construction can be understood by considering a system of D$p$-D$(p+4)$ branes  \cite{Douglas:1995bn,Douglas:1996uz,Witten:1994tz}. For $U(r)$ gauge group with $n$ instantons, we consider $n$ D$p$-branes and $r$ D$(p+4)-$branes. From the point of view of the D$(p+4)$-branes, the D$p$-branes are thought of as $n$ instantons of the $U(r)$ gauge theory when they are on the top of the $p+4$-dimensional branes. From the point of view of the world-volume theory on the D$p$-branes, this is the case when the theory is in the Higgs branch. Note that therefore one can relate the Higgs branch of the world-volume theory on the D$p$-branes to the (centered) moduli space of $n$ instantons with gauge group $U(r)$, an observation we will return to later. Note also that the from world-volume theory on instantons, the gauge symmetry of the spacetime appears as a global symmetry.
It is possible to generalize this construction to other classical gauge groups $Sp/SO$ by adding an orientifold plane to above configuration. Then we get either $O(n)$ gauge group on instantons or $Sp(n/2)$ gauge groups. The corresponding global symmetry groups are $Sp(r)$ for the former and $O(2r)$ or $O(2r+1)$ for the latter. Here the global symmetry is actually the gauge group from the point of view of the spacetime. 

There are similar constructions for the $E_{6, 7, 8}$ groups. One possible construction can be obtained from F-theory \cite{Sen:1996vd, Banks:1996nj}. In F-theory, there are 7-branes with exceptional gauge symmetries on it. We probe such geometry using $n$ D3-branes. The number $n$ is called the rank of the theory and it is the same as the complex dimension of the Coulomb branch. The D3-brane worldvolume theory now has exceptional global symmetry arising from D3-D7 strings. Now it is natural to expect the Higgs branch of this theory is the moduli space of the exceptional instantons. This theory admits alternative description in terms of multiple M5-branes wrapped on a 3-punctured sphere  \cite{Gaiotto:2009we}. When $n=1$, this is the Minahan-Nemeschansky theory \cite{Minahan:1996fg,Minahan:1996cj}, which for $E_6$ is also called the $T_3$ theory. The theories that correspond to higher $n$ are proposed by \cite{Benini:2009gi}.

Unfortunately these constructions for exceptional Lie groups do not lead to a construction of the
instanton moduli space itself. The only case where the moduli space is known is
for one instanton, where the partition function takes a
very simple and uniform form \cite{Gaiotto:2008nz,VinbergPopov, Benvenuti:2010pq, Keller:2011ek}. 
The equivariance group of the problem is given by
\be \label{equivariancegroup}
G\times U(1)_{\e_1}\times U(1)_{\e_2}\ .
\ee
It is useful to consider the perform the counting in five dimensions.
In this case the instanton partition function becomes
the Hilbert series, that is the character of the equivariance
group on the holomorphic functions on $\CM_k$. It is often
convenient to take the diagonal subgroup $U(1)_\tau$
and its orthogonal $U(1)_x$, so that $\tau =e^{\half(\e_1+\e_2)}$
and $x=e^{\half(\e_1-\e_2)}$. The one instanton term is then
\be\label{inst5dk1}
Z_1= \frac{1}{(1-e^{\e_1})(1-e^{\e_2})}\sum_{k\geq0}
\chi_{k\theta}(a_i)\tau^{2k}\ .
\ee
Here $\theta$ is the highest root of $G$, and
$\chi_{k\theta}$ is the character of the representation
$k\theta$.

Even though we do not know the moduli space for multiple
instantons, the simple and above all universal form of
(\ref{inst5dk1}) implies that it may be possible at least
to construct their Hilbert series in terms of
representations of the equivariance group (\ref{equivariancegroup}). 
One obvious approach is to simply take the result
of the classical Lie groups and try to rewrite it
in a universal way, in this way extrapolating it
to exceptional Lie groups.
To put it another way, once we can express
the result terms of properties of the group $G$ only,
it will be obvious to generalize to arbitrary groups.

One way to do this is to use results derived by Nakajima and Yoshioka.
In a series of papers
\cite{Nakajima:2003pg, Nakajima:2005fg, Nakajima:2003uh} they
consider $U(r)$ instantons on the blowup of $\IR^4 = \IC^2$.
The instanton counting on the blowup can be related
to instantons on ordinary $\IC^2$. This allows 
to derive recursion relations in the instanton number $n$.
These recursion relations are a generalization
of the contact term equations for the Seiberg-Witten prepotential
discussed in \cite{Losev:1997wp,Losev:1997tp, Gorsky:1998rp}.
Their virtue is that they only 
contain purely group theoretic data of $U(r)$.
It is thus immediately obvious how they can
be extrapolated to arbitrary gauge groups.
This allows to compute the instanton partition function
for an arbitrary gauge group at arbitrary instanton number.
 
As mentioned above, in some cases the moduli space of instantons 
corresponds to the Higgs branch of a certain other theory.
In some of those cases it is possible to compute the Hilbert series of the Higgs branch,
and so compare it to our results.
In particular, using the TQFT structure of the superconformal index for the theories coming from the M5-branes (called the theories of class $\CS$) \cite{Gadde:2009kb} one can write down a conjecturual expression for the index of rank $n$ $E_r$ theories \cite{Gadde:2010te,Gadde:2011uv}. Moreover \cite{Gadde:2011uv} observed that the specialization of the index to the so-called Hall-Littlewood index is equivalent to the Hilbert series on the Higgs branch, which
establishes the connection to the instanton partition function.\footnote{The relation between the superconformal index and the Hilbert series of the moduli space of $\CN=1$ theories was first observed in \cite{Spiridonov:2009za}.}

In this paper we show that the extrapolation of the results of Nakajima and Yoshioka to
gauge groups other than $U(r)$ indeed works. In particular
we show that for 1 instantons it agrees with the partition function derived in \cite{Keller:2011ek}. 
We also show that for 2 instantons, the result agrees with the instanton counting 
for classical Lie groups when applicable, and that for $E_6$
and $E_7$ it agrees with the index of the corresponding
strongly coupled theories with flavor
symmetry $E_6$ and $E_7$ computed in \cite{Gadde:2010te,Gadde:2011uv,Gaiotto:2012uq}.

\vspace{.4cm}
\noindent
\emph{Note added:} We coordinated publication with \cite{Hanany:2012dm}, which computes the Hilbert series of the two instanton moduli space using methods different from ours.

\section{Recursion relations from the blowup}

In this section, we summarize the method of Nakajima-Yoshioka \cite{Nakajima:2003pg, Nakajima:2005fg, Nakajima:2003uh} to compute the instanton partition function for arbitrary gauge group without matter fields. 

The blowup equation enables us to determine the instanton partition function recursively in the instanton number. Schematically, the idea of Najakima-Yoshioka is as follows. We first compute the instanton partition function $\hat{Z}$ on the blow up $\hat{\IC}^2$ of $\IC^2$, which is closely related to $Z$, the partition function on $\IC^2$.\footnote{Note that normally one uses known partition functions $Z$ on $\IC^2$ as an input to the blowup equation to compute $\hat Z$. (See for example \cite{Bonelli:2011jx, Bonelli:2011kv} for such a computation on ALE spaces.) Here of course
we are using the blowup equation the other way around, that is to obtain expressions for $Z$ from $\hat Z$. } 
More generally, denoting by $C$ the exceptional divisor of the blowup, we can compute the correlation
function $\hat{Z}_d$ of $\mu(C)^d$ on $\hat{\IC}^2$, where $\mu(C)$ is constructed in a specific way.
Schematically, $\hat{Z}_d$ is related to $Z$ by
\be
 \hat{Z}_d \sim \sum_{\vec{k}} q^{\half (\vec{k}, \vec{k}) } f_d(\vec{k}) Z^{(1)}(\vec{k}) Z^{(2)}(\vec{k}) 
\ee 
where $Z^{(i)}$ is $Z$ with a shift in some variables, and $f_d(\vec{k})$ is some function. 
$\mu(C)$ is constructed in such a way that one can apply vanishing theorems to show that
in fact $\hat{Z}_d=Z$ for $d=0,1\ldots,r-1$.
Using this we can obtain enough equations to determine $Z$ recursively in the instanton
number from $f_d(\vec{k})$.
The central observation is that $f_d(\vec{k})$ is given
in terms of group theoretic data of the gauge group $G$, so that we can compute
$Z$ even if we do not have an explicit construction of the instanton moduli space.

The limiting behavior of the blowup equation can be used to show that the leading prepotential part
\be
F_0 = \lim_{\e_{1, 2} \to 0} \e_1 \e_2 \log (Z_{Nek}) 
\ee
of the Nekrasov partition function satisfy the contact-term equation \cite{Losev:1997wp,Losev:1997tp, Gorsky:1998rp}. The contact term equation is known to be satisfied by the Seiberg-Witten prepotential and has a unique solution \cite{Edelstein:1998sp}. Therefore they prove that the $F_0$ part of the Nekrasov partition function indeed reproduces the Seiberg-Witten prepotential.

\subsection{Instanton counting in 4d and 5d}
Let us first briefly recapitulate counting instantons of $G=U(r)$ gauge theory on $\IC^2$. The naive moduli space of $G-$instantons is singular. To cure this issue we deform the instantons to non-commutative instantons. This new moduli space of $U(r)$-instantons is known to be identical to the moduli space of framed torsion-free sheaves $E$ on $\IP^2$ with rank $r$. 
The instanton number of $E$ is given by its second Chern number $c_2(E)$, and we will denote its moduli space
as $M(r, n)$. On this space there is a natural action of the Cartan torus of $SO(4)$, as well as of the group of
large gauge transformations $G$, which change the framing. We will denote the total torus of these groups by $T = U(1)^2_{\e_1, \e_2} \times U(1)^r$. To cure the IR divergence of the integral over the moduli space, we put the gauge theory on $\Omega$-background, that is we consider the equivariant integral with respect to the tours action $T$. 

The Nekrasov partition function is defined as
\be
 Z(\e_1, \e_2, \vec{a}) = \sum_{n=0}^{\infty} q^n \oint_{M(r, n)} 1
\ee
where the integral $\oint 1 $ is the equivariant volume
\be
 \oint_{M(r, n)} 1 = \int_{M(r, n)} \exp \big(\omega + \mu_T (\vec{a}, \e_1, \e_2) \big)
\ee
where $\omega$ is a symplectic form on $M(r, n)$ and $\mu_T$ is the moment map corresponding to the rotation of $T$. 
This integral localizes to the fixed points of $T$ and therefore only depends on the tangent space
at the fixed points,
\be
 Z(\e_1, \e_2, \vec{a}) = \sum_{\vec{Y}}  \frac{q^{|\vec{Y}|}}{e(T_{\vec{Y}})} \ ,
 \ee 
where $e(T_{\vec{Y}})$ is the equivariant Euler class of the tangent bundle of $M(r, n)$ at the fixed point $\vec{Y}$. The fixed points of $M(r, n)$ are labeled by $r-$colored Young diagrams $\vec{Y}$ with total number of boxes $n$.\footnote{The localization onto the $T-$fixed points seems to be much more subtle in the case of $SO/Sp$ instantons. See appendix B of \cite{Hollands:2010xa} for a discussion.} 
Note in particular that for instanton counting on $\IC^2$, the action of 
the $\Omega$ deformation group localizes the integral at $z_1=z_2=0$. The Euler class of the tangent bundle at the fixed point 
\be
e(T_{\vec{Y}})
 =\displaystyle \prod_{\a, \b} n^{\vec{Y}}_{\a, \b} (\e_1, \e_2, \vec{a})  , 
\ee 
can be written in a closed form $n^{\vec{Y}}_{\a, \b}$ as the product
of factors over hook-lengths of $\vec{Y}$.

Counting instantons in 5d is very similar.
In fact it is often simpler to perform the counting in 5d and then
obtain the 4d result by taking an appropriate $\b\rightarrow0$ limit \cite{Nekrasov:2004vw}.
The 5d or $K-$theoretic partition function for pure $U(r)$ $\CN=2$ SYM is defined as follows:
\be
 Z(\e_1, \e_2, \vec{a}; q, \beta) = \sum_{n=0}^\infty \left( q \b^{2r} e^{-r \b (\e_1 + \e_2)/2} \right)^n Z_n (\e_1, \e_2, \vec{a} ; \beta)
\ee
where
\be
Z_n (\e_1, \e_2, \vec{a}; \beta) =  \sum_{i=0}^\infty (-1)^i \textrm{ch} H^i (M(r, n), \CO) =  \textrm{ch} H^0 (M(r, n), \CO) . 
\ee
Here we evaluate the character of the torus action $T = U(1)^2 \times U(1)^r$. Since the higher cohomology vanishes, we simply need to evaluate the character over $H^0 (M(r, n), \CO)$,\footnote{Actually, one can show that it is enough to compute the character of the holomorphic function on the singular algebraic variety $M_0(r, n)$ constructed from ADHM, not the resolved space $M(r, n)$.} which is the ring of holomorphic functions on the moduli space of instantons on $\IC^2$. From \cite{Nakajima:2003pg, Nakajima:2005fg}, we get
\be
 Z(\e_1, \e_2, \vec{a}; q, \beta) = \sum_{\vec{Y}} \frac{(q \beta^{2r} e^{-r \beta (\e_1 + \e_2)/2})^{|\vec{Y}|}}{\prod_{\a, \b} n^{\vec{Y}}_{\a, \b} (\e_1, \e_2, \vec{a}; \b)}
\ee 
where $n^{\vec{Y}}_{\a, \b} (\e_1, \e_2, \vec{a}; \b)$ is also some function similar to the 4d case, which reduces to the 4d result when we take $\b \to 0$ limit. 

\subsection{Instanton counting on the blowup and the recursion relation}

Let us now turn to the moduli space on $\IC^2$ blown up at the origin. This blowup $\hat{\IC^2}$ 
is defined as a submanifold of $\IC^2 \times \IP^1 $ defined by 
\be
\{ (x, y), [z: w] \in \IC^2 \times \IP^1 | xw = yz \} . 
\ee
Here $[z: w] \in \IP$ are the homogeneous coordinates. The projection $p$ projects to the first factor. The set $C \subset \hat{\IC^2}$ with $x=y=0$ is called the exceptional divisor. 

As in the previous case, we consider the framed moduli space $\hat{M}(r, k, n)$ of torsion-free sheaves on $\hat{\IP}^2$ with rank r, with instanton number $n$. A new ingredient is that we can have non-vanishing first Chern class on $C$ with $\langle c_1(E),[C]\rangle=k$ which measures magnetic flux through the compact 2-cycle $C$. The moduli space $\hat{M}(r, k, n)$ is again the appropriate compactification of the moduli space of instantons on $\hat{\IC}^2$. 

The first crucial ingredient in deriving the recursion relations is
the observation that the instanton partition function on $\hat{\IC}^2$ is
related to the one on $\IC^2$.
More precisely,
letting $U(1)^2_{\e_1, \e_2}$ act as
\be
((x, y), [z: w]) \mapsto ((e^{i\e_1}x, e^{i\e_2}y),[e^{i\e_1}z:e^{i\e_2}w])\ ,
\ee
Nakajima-Yoshioka prove that the fixed points of $\hat{M}(r, k, n)$ are labeled by two colored Young diagrams and
an $r-$dimensional vector $(\vec{k}, \vec{Y}^1, \vec{Y}^2)$. The two Young diagrams label the two fixed points
at $((0,0),[0:1])$ and $((0,0),[1:0])$ respectively, and $\vec{k}$ comes from the fact that $C$ as a whole is
invariant under $U(1)^2_{\e_1, \e_2}$. The Euler class of the tangent space at the fixed point then
factorizes into three components,
\be
 \prod_{\a, \b} l^{\vec{k}}_{\a, \b} (\e_1, \e_2, \vec{a}) ~ n^{\vec{Y_1}}_{\a, \b}(\e_1, \e_2 - \e_1, \vec{a} + \e_1 \vec{k} ) ~ n^{\vec{Y_2}}_{\a, \b} (\e_1 - \e_2, \e_2, \vec{a} + \e_2 \vec{k} ) , 
\ee
where the contributions from $\vec{Y}^1$ and $\vec{Y}^2$ are the same as for the fixed
points on $\IC^2$, up to some shifts in the parameters.
The function $l^{\vec{k}}_{\a, \b} (\e_1, \e_2, \vec{a})$
will be discussed in a moment. It is the analog of the contact
term in the contact-term equations of \cite{Losev:1997wp,Losev:1997tp}.
Let us just note that this means that we can express
the instanton partition function $\hat Z$ on the blow up
in terms of
the partition function on $\IC^2$ as
\be\label{hatZexpr}
 \hat{Z} (\e_1, \e_2, \vec{a}; q, \beta) &=& \sum_{\vec{k}} \frac{\left( e^{-\b (\e_1 + \e_2) r/2} q \b^{2r} \right)^{(\vec{k}, \vec{k})/2} }{\prod_{\a,\b} l^{\vec{k}}_{\a,\b} (\e_1, \e_2, \vec{a}) } \times  \\
 &{ }& ~~ Z (\e_1, \e_2 - \e_1, \vec{a}+\e_1 \vec{k}; e^{-\b \e_1 r/2}q, \b) Z (\e_1-\e_2, \e_2, \vec{a} + \e_2 \vec{k}; e^{-\b \e_2r/2} q, \b) . \nn 
\ee
To obtain the second ingredient for the recursion relation,
let us generalize the instanton partition function.
Let us consider a line bundle $\CO(C)$ associated to the divisor $C$ and its $d$-th tensor power $\CO(dC)$. 
Define the 4d version of the correlation function as
\be
\hat{Z}_d^{4d} = \sum_{n=0}^{\infty} q^n \oint_{M(r, n)} \mu(C)^d\ ,
\ee
where $\mu(C) \in H^*_T (\hat{M}(r, k, n))$ is a certain equivariant cohomology class defined
in \cite{Nakajima:2003pg, Nakajima:2005fg, Nakajima:2003uh}.
One can define 5d version of the correlation function as well:
\be
 \hat{Z}_d = \sum_{n=0}^\infty \left(q \b^{2r} e^{-r\b (\e_1 + \e_2)/2}\right)^n \textrm{ch} H^0 \left( M(r, n), \CO(d C) \right) . 
\ee
Similar as when we introduce line bundles for theories
with hypermultiplets, this leads to the same expression
as (\ref{hatZexpr}) for $\hat{Z}_d$, but with an additional
factor to the power of $d$ in the numerator coming from the localization of the $\mu(C)$ at the fixed point.
The exact expression for $\hat{Z}_d$ is given by (\ref{blowupeq5d})
in the next section.
The point is that $\mu(C)$ was constructed in such a way that $\hat{Z}_d$
is simply the ordinary instanton partition function,
\be \label{vanish}
\hat Z_d(\e_1, \e_2, \vec{a}; q, \beta)=Z(\e_1, \e_2, \vec{a}; q, \beta)\qquad 0\leq d < r
\ee
when the magnetic flux $k$ through $C$ is turned off. We can now use (\ref{vanish}) and the generalization of (\ref{hatZexpr}) for $\hat{Z}_d$
to derive $Z$ recursively order by order in $q$.
To do this we take the equations for $d=0,1,2$, which gives three equations for
the three unknowns: $Z$, and the two $Z$ with shifted
arguments that appear on the right hand side of (\ref{hatZexpr}). At 
a given order in $q$, these equations can then be used to determine
$Z_n$ in terms of the lower $Z_k, k<n$. The only input necessary
for this procedure are the functions $l^{\vec{k}}_{\a, \b} (\e_1, \e_2, \vec{a})$.

For gauge group $U(r)$, the expressions for the $l^{\vec{k}}_{\a, \b} (\e_1, \e_2, \vec{a})$
were found in \cite{Nakajima:2005fg}. In fact they can be written
in terms of group theoretic data of $U(r)$: The product over
$\a,\b$ becomes a product over roots $\vec{a}\in\Delta$, and
the $\vec{k}$ is summed over all vectors in the lattice
spanned by the coroot lattice.

Even though all of this was only shown for gauge group $U(r)$,
it is then immediately obvious what the conjecture for the partition function
of a general gauge group $G$ group is. We simply define $l^{\vec{k}}_{\vec{a}}$
in terms of the roots and the coroot lattice of $G$, and then use again
relations (\ref{vanish}) and (\ref{hatZexpr}) to compute the 
coefficients of $Z$ recursively. In the rest of this article,
we will check this conjecture against various other computations.
Let us now give the explicit recursion relations for arbitrary $G$.

\subsection{Recursion relations for arbitrary $G$}
Let $G$ be a Lie group of rank $r$. Let $h^\vee$ be its
dual Coxeter number, $\Delta$ its set of roots, and $\Delta_l$
the set of long roots. In our convention a long root $\gamma$
has norm squared 2, $(\gamma,\gamma)=2$ where $(~,~)$ is the standard inner product 
the weight space of $G$. By a slight abuse of notation we will always identify
the weight space of $G$ with its dual.
Let $\a_i, i=1,\ldots, r$
be the simple roots, and $\a_i^\vee$ the corresponding 
coroots $\a_i^\vee = \frac{2\a_i}{(\a_i, \a_i)}$. 
Finally denote by $K^\vee$ the coroot lattice
spanned by vectors $\vec{k}=\sum_i k^i\a_i^\vee , k^i \in \IZ$.

From these quantities we can then define the functions
\be
 l^{\vec{k}}_{\a} (\e_1, \e_2, \vec{a}) 
  = \begin{cases}
  {\displaystyle \prod_{\substack{i, j \ge 0 \\ i+j \le -(\vec{k}, \a ) - 1}}} \left(1- e^{\beta \left(i \e_1 + j \e_2 -  ( \vec{a}, \a ) \right)} \right) & \textrm{if $( \vec{k}, \a ) < 0$} \\ {\displaystyle \prod_{\substack{i, j \ge 0 \\  i+j \le ( \vec{k}, \a ) - 2}}} \left(1-e^{\beta \left( -(i+1) \e_1 - (j+1) \e_2 -  ( \vec{a}, \a ) \right)} \right) & \textrm{if $( \vec{k}, \a ) > 1$} \\
  1 & \textrm{otherwise}  \end{cases} .
\ee

Now, according to the conjecture, the instanton partition function on the blow up is 
still related to the standard instanton partition function by
\be \label{blowupeq5d}
 \hat{Z}_{d} (\e_1, \e_2, \vec{a}; q, \beta) &=& \sum_{\vec{k}\in K^\vee} \frac{\left( e^{\b (\e_1 + \e_2) (d-h^\vee/2)} q \b^{2h^\vee} \right)^{(\vec{k}, \vec{k})/2} e^{\b (\vec{k}, \vec{a}) d}}{\prod_{\vec{\a} \in \Delta} l^{\vec{k}}_{\vec{\a}} (\e_1, \e_2, \vec{a}) } \times  \\
 &{ }& ~~ Z (\e_1, \e_2 - \e_1, \vec{a}+\e_1 \vec{k}; e^{\b \e_1 (d-h^\vee/2)}q, \b) Z (\e_1-\e_2, \e_2, \vec{a} + \e_2 \vec{k}; e^{\b \e_2 (d-h^\vee/2)} q, \b) . \nn 
\ee
Moreover, we assume that (\ref{vanish}) still holds.
Combining those two relations and expanding order by order in the instanton number, we get the following relation:
\be \label{blowupeq5d2}
 Z_{n} (\e_1, \e_2, \vec{a}; \b) &=& \sum_{\half (\vec{k}, \vec{k}) + l + m = n} 
 \exp \left[ \b d \left( l \e_1 + m \e_2 + (\vec{k}, \vec{a}) + \frac{(\vec{k}, \vec{k})}{2} (\e_1 + \e_2) \right) \right] \frac{1}{\prod_{\a \in \Delta} l^{\vec{k}}_{\a} (\e_1, \e_2, \vec{a}) }\nn  \\
&{ }&~~~~~~~~~~~\times Z_l (\e_1, \e_2 - \e_1, \vec{a} + \e_1 \vec{k}; \b) Z_m (\e_1 - \e_2, \e_2, \vec{a} + \e_2 \vec{k}; \b) .   
\ee
This is the key formula. 
To solve it recursively, let us write
\begin{multline}
Z_{n} (\e_1, \e_2, \vec{a}; \b) = e^{\beta\e_1dn}Z_n(\e_1, \e_2-\e_1, \vec{a}; \b)
+ \\e^{\beta\e_2 d n}Z_n(\e_1-\e_2, \e_2, \vec{a}; \b)+I_n^{(d)}\ , \qquad d=0,1,\ldots r-1\ ,
\end{multline}
where we have defined
\be 
 I_{n}^{(d)}(\e_1, \e_2, \vec{a}; \b) &=& \sum_{\displaystyle \substack{\half (\vec{k}, \vec{k}) + l + m = n \\ l,m < n}}
\frac{\exp \left[\beta d \left( l \e_1 + m \e_2 + (\vec{k}, \vec{a}) + \frac{(\vec{k}, \vec{k})}{2} (\e_1+\e_2) \right)\right]}{\prod_{\a \in \Delta} l^{\vec{k}}_{\a} (\e_1, \e_2, \vec{a}) }  \\
 &{ }& ~~~~~~~~~~~~~~~~ \times Z_l (\e_1, \e_2 - \e_1, \vec{a} + \e_1 \vec{k}; \b) Z_m (\e_1 - \e_2, \e_2, \vec{a} + \e_2 \vec{k}; \b)  \nn   
\ee
so that it only depends on $Z_{m,l}$ with $l,m<n$.
Using the above equation for $d=0, 1, 2$ we can solve for $Z_n (\e_1, \e_2, \vec{a};\b)$ to get
\be \label{5dreeq} 
 Z_n (\e_1, \e_2, \vec{a}; \b)= \frac{{ e^{\b n(\e_1+\e_2)} } I_{n}^{(0)} - (e^{\b n \e_1} + e^{\b n \e_2}) I_{n}^{(1)} + I_{n}^{(2)}}{(1-e^{\b n\e_1} )(1-e^{\b n\e_2})} \ . 
\ee
Since $I_n^{(d)}$ is determined completely by instanton partition function of instanton number lower than $n$, we get the partition function at any $n$ by recursively using \eqref{5dreeq} and $Z_0(\e_1, \e_2, \vec{a}; \b) = 1$. 
One can get the 4d version of the formula by simply taking the $\beta \to 0$ limit with appropriate scaling. 


\section{Instanton partition functions for exceptional groups}
In this section, we discuss various applications of the formula \eqref{5dreeq}. First, we derive the 1-instanton formula of \cite{Keller:2011ek} for arbitrary gauge groups. We also compute 2-instanton partition function and propose a general expression up to a certain order in $\tau = e^{\half \b(\e_1 + \e_2)}$ expansion. Finally, we check our computation for $E_{6, 7}$ against the exceptional indices computed in \cite{Gaiotto:2012uq}. 

\subsection{One instanton formula} \label{oneinst}
As a warmup, let us reproduce the one instanton expression (\ref{inst5dk1}) from (\ref{blowupeq5d}). 
To simplify notations we will absorb $\beta$ from now on, and write $\a$ instead of $(\a,\vec{a} )$.
For $\vec{k}\in K^\vee$ define $L(\vec k) = \prod_{\a\in\Delta} l^{\vec k}_{\a}(\e_1,\e_2,\vec a)$.

The sum in $I^{(d)}_1$ is then only over
only vectors of length squared 2. The only such vectors
are given by the coroots $\gamma^\vee$ of long roots $\gamma$. Note that in our conventions we have
$\gamma=\gamma^\vee$ for long roots.
For these $\gamma$ we get
\be
L(\gamma)= (1-e^{-(\e_1+\e_2+\gamma)})(1-e^{\gamma})(1-e^{(\e_1+\gamma)})(1-e^{(\e_2+\gamma)})
\prod_{ (\a, \gamma) =1}(1-e^{\a})\ ,
\ee
which gives
\be
I^{(d)}_1=e^{(\e_1+\e_2)d}\sum_{\gamma\in \Delta_l} \frac{e^{\gamma d}}{L(\gamma)}
\ee
Using (\ref{5dreeq}) then gives 
\begin{multline}
Z_1= \frac{e^{\e_1+\e_2} I^{(0)}_1-(e^{\e_1}+e^{\e_2})I^{(1)}_1+I^{(2)}_1}{\left(1-e^{\e_1}\right) \left(1-e^{\e_2}\right)}
= \frac{e^{\e_1+\e_2}}{(1-e^{\e_1})(1-e^{\e_2})}\sum_{\gamma\in \Delta_l}\frac{(1-e^{\gamma+\e_2})(1-e^{\gamma+\e_1})}
{L(\gamma)}\\
= \frac{e^{(\e_1+\e_2)}}{(1-e^{\e_1})(1-e^{\e_2})}\sum_{\gamma\in \Delta_l}\frac{1}
{(1-e^{-\e_1-\e_2+\gamma})(1-e^{-\gamma})
\prod_{\a\cdot\gamma =1}(1-e^{-\a})}\ ,
\end{multline}
where in the last line we have relabeled the dummy variable $\gamma\rightarrow-\gamma$.
To get agreement with \cite{Keller:2011ek}, we apply the identity
\be
e^{\gamma/2}\prod_{(\a, \gamma) =1} e^{\a/2}= e^{(h^\vee-1)\gamma/2}\ ,
\ee
which gives 
\be
Z_1=\frac{1}{(1-e^{-\e_1})(1-e^{-\e_2})}\sum_{\gamma\in \Delta_l}\frac{e^{(h^\vee-1)\gamma/2}}
{(1-e^{-\e_1-\e_2+\gamma})(e^{\gamma/2}-e^{-\gamma/2})
\prod_{\a\cdot\gamma =1}(e^{\a/2}-e^{-\a/2})}\ .
\ee 
This agrees indeed with (B.3) in \cite{Keller:2011ek} after redefining $\e_{1,2}\rightarrow-\e_{1,2}$.

Note that $Z_1$ written as a rational function in $\tau=e^{\half( \e_1+\e_2) }$ is palindromic,
that is $Z_1(1/\tau)=\tau^{2h^\vee}Z_1(\tau)$, which means that the coefficients
of $\tau^k$ of the numerator and denominator polynomials are symmetric under 
$a_k\leftrightarrow a_{N-k}$, where $N$ is the degree of the polynomial. As is discussed
in \cite{Forcella:2008bb,Hanany:2008kn}, using 
a theorem due to Stanley \cite{Stanley197857} this general property
follows from the fact that the moduli space is given by a hyperk\"ahler affine
variety.
From palindromy it follows that replacing $\e_{1,2}\rightarrow-\e_{1,2}$ 
is equivalent to multiplying by an overall factor $e^{h^\vee(\e_1+\e_2)} = \tau^{2h^\vee}$.
In the following we will always suppress this factor. 

\subsection{Higher instantons}
The one instanton partition function depended on $U(1)_x$
only in a trivial way, that is through the contribution
of the center of mass. For two and more instantons,
the $U(1)_x$ dependence is more complicated, since now
the relative positions of the instantons matter. 
Since all represenations of $U(1)_x$ have integer charge,
and since the partition function is invariant under $x\mapsto x^{-1}$,
it turns out that its representations actually enhance
to $SU(2)$ represenations. This is the additional $SU(2)_x$
that was observed in \cite{Gaiotto:2012uq}.

In general we can always obtain a closed expression for $Z_k$
by applying (\ref{5dreeq}) repeatedly. Unfortunately for
higher rank gauge groups and higher instanton numbers this
expression becomes very unwieldy. On general grounds we know
that $Z_k$ is a representation of the equivariance group
$U(1)_\tau\times U(1)_x \times G$, but its representation
structure is not immediately obvious. Nonetheless, it
is in principle straightforward to write this expression
in terms of characters. From the form of (\ref{5dreeq}) it is clear
that 
\be
Z_k = \frac{P_k(x,\tau,\vec{a})}{Q_k(x,\tau,\vec{a})}
\ee
can be written as a rational function of two polynomials.
In fact, in the cases we checked, $P$ and $Q$ turn out to
be palindromic.
We can then decompose $P$ and $Q$ into representations
of the equivariance group by comparing a finite number
of coefficients each. This gives a closed expression
in terms of characters of the equivariance group.

To make the previous discussions more concrete, let us discuss the case 
of $G_2$ in more detail. Let us denote by $\chi_k$ the $k$ dimensional
representation of $SU(2)_x$.
We can express $Z_1$ in this way with
\be
P_1= (1 + \tau^2 + \tau^4 + \tau^6) + (\tau^2 + \tau^4) \chi^{G_2}_{[1,0]}
\ee
and
\begin{multline}
Q_1=(1 + \tau^2 - \tau \chi_2) \Big( (1 - \tau^2 + \tau^4) (1 + \tau^2 + \tau^4)^2 - 
   \tau^2 \big[ (1 + \tau^4)^2 \chi^{G_2}_{[0, 1]} + \\
      \tau^4 \chi^{G_2}_{[0, 2]} - (1 + \tau^4 + \tau^8) \chi^{G_2}_{[1, 0]} + 
      \tau^2 (1 + \tau^4) \chi^{G_2}_{[1, 1]} \big] + 
   \tau^4 (1 + \tau^2 + \tau^4) \chi^{G_2}_{[3, 0]} \Big)\ ,
\end{multline}
where we have labeled the representations of $G_2$
by the usual Dynkin labels. With the
adjoint representation given by $\chi^{G_2}_\theta=\chi^{G_2}_{[0,1]}$
it is straightforward to check that this agrees with (\ref{inst5dk1}).

For $Z_2$ the expressions for $P_2$ and $Q_2$ are more complicated. For
brevity we will only give
the unrefined expressions, from which the dimensions of the representations
can be read off:
\be
P_2 &=&(1 + 9 \tau^2 + 44 \tau^4 + 86 \tau^6 + 2 \tau^8 - 215 \tau^{10} + 30 \tau^{12} + 737 \tau^{14} + 591 \tau^{16} - 529 \tau^{18} + \ldots + \tau^{38}) \nonumber \\
 &{ }&~+ \tau^3(8 + 30 \tau^2 + 16 \tau^4 - 222 \tau^6 - 327 \tau^8 + 150 \tau^{10} + 689 \tau^{12}  - 340 \tau^{14}  - 1088 \tau^{16} + \ldots + 8 \tau^{32})\chi_2 \nonumber \\
  &{ }&~+ \tau^6(8 - 33 \tau^2 - 117 \tau^4 - 54 \tau^6 + 331 \tau^8 + 234 \tau^{10} - 369 \tau^{12} +\ldots + 
 8 \tau^{26})\chi_3  \\
 &{ }&~+ \tau^9(1 - 6 \tau^2 + 64 \tau^4 + 112 \tau^6 + 14 \tau^8 - 118 \tau^{10} + \ldots + \tau^{20})\chi_4 \nonumber \\
 &{ }&~-\tau^{16}(7 + 29 \tau^2 + 29 \tau^4 + 7 \tau^6)\chi_5  \nonumber
\ee
and
\be
Q_2=(1-\tau^2)^6(1-\tau^3\chi_2+\tau^6)^6(1-\tau\chi_2+\tau^2)^2(1+\tau\chi_2+\tau^2)\ .
\ee
The expressions are again palindromic, so that for simplicity we have omitted terms that are fixed by palindromy.
This suggests that the higher instanton moduli space for $G_2$ is still a hyperk\"ahler affine variety.

\subsection{General 2-instanton partition function}
As seen above, the expressions involved become 
very complicated. Still, one may hope to find a
universal expression for $Z_2$ in the spirit of (\ref{inst5dk1}).
For the first few terms in $\tau$ we have actually
managed to do that.

For an arbitrary Lie group, denote by $\chi_{k\theta}$ the 
representation of Dynkin label $\theta_k = k\theta$,
where $\theta$ is the highest root so that $\chi_\theta$ is the 
adjoint representation.
As a generalization of \cite{Gaiotto:2012uq}, 
we claim that in general the $k=2$ instanton term
can be written as
\begin{multline}\label{Z2general}
Z_2= \frac{1}{(1-x \tau)(1-x^{-1} \tau)}\Big(
1 + (\chi_{\theta} + \chi_3)\tau^2 + \chi_{\theta}\chi_2 \tau^3\\
 +(\chi_{\theta}\chi_3+\chi_{\Sym^2\theta}+\chi_{5})\tau^4
+\left(\chi_{\theta}\chi_4+(\chi_{2\theta}+\chi_{\Alt^2 \theta})\chi_2 \right) \tau^5 +\ldots \Big) 
\end{multline}
where $\Sym^n$ and $\Alt^n$ denote the $n$th symmetric and alternating
power respectively. It is certainly intriguing that up to this order it is possible
to find such an expression involving only positive multiplicities
of symmetric and alternating powers of $\chi_{k\theta}$. 
In particular note that this is a homogeneous expression
if we assign grades $[\chi_n]=n-1$, $[\chi_{\theta}]=2$ and
$[\tau]=-1$. We have checked that this relation holds for $G=A_{1, 2, 3, 4}, B_{2, 3, 4}, C_{3, 4}, D_4, E_{6, 7}, F_4, G_2$. 

At order $\tau^6$ we obtain  
\be
 \left(\chi_7+\chi_5\chi_{\theta}+\chi_3(\chi_{\Sym^2\theta}+\chi_{2\theta})+\chi_{\Sym^3\theta}  - C(G) \right) \tau^6 . 
\ee
Unfortunately we have not been able to express the last term $C(G)$ in 
a similar universal way.
For example, we find the $C(G)$ to be
\be
 C(SU(2)) &=& 0 ,\\
 C(SU(3)) &=& \chi^{SU(3)}_{[0, 0]} = 1 , \\
 C(SU(4)) &=& \chi^{SU(4)}_{[1, 0, 1]} + \chi^{SU(4)}_{[0, 0, 0]} , \\
 C(SU(5)) &=&  \chi^{SU(5)}_{[0, 1, 1, 0]} + \chi^{SU(5)}_{[0, 0, 0, 0] },
\ee
where the representations are denoted using the standard Dynkin label notation.

\subsection{Comparison with the exceptional index}
Let us compare the instanton partition function obtained from the recursion relation with the result from the superconformal indices recently proposed in \cite{Gaiotto:2012uq}. 
This superconformal index is related to the instanton partition function in the following way.
First, realize a superconformal field theory with $E_r$ flavor symmetry in terms of multiple M5-branes wrapped on a 3-punctured sphere with non-maximal flavor symmetry, which enhances to $E_r$. The complex dimension of the Couloumb branch of this theory is called its rank. 
Let us denote the theory with flavor symmetry $E_r$ and rank $n$ by $E_{r, n}$.
For example, $E_{6,2}$ can be realized by 6 M5-branes wrapped on a 3-punctured sphere with 3 $SU(3)$ punctures.  The Higgs branch of the $E_{r, n}$ theory is then the moduli space of instantons $\CM^{E_r}_{n}$ with gauge group $E_r$ and instanton number $n$ 
\be
 \CM^{E_r}_{n} = \textrm{Higgs}(E_{r, n})\ . 
\ee
To obtain the instanton partition function we thus need to compute the Hilbert series of the Higgs branch.
In \cite{Gadde:2011uv} it was argued that for theories coming from genus 0 UV-curves 
this Hilbert series is given by the superconformal index (more precisely, the Hall-Littlewood index).
At the same time they gave a recipe to compute this index for strongly coupled theories and
checked that for $E_{6, 1}$ it indeed agrees with (\ref{inst5dk1}).
Unfortunately their computation method breaks down for theories of higher rank,
as it give divergent expressions for the coefficients of the $\tau$ expansion.
Recently \cite{Gaiotto:2012uq} proposed that
this divergence comes from a decoupled hypermultiplet in the theory. They find a prescription for rendering the index finite that involves assigning a fugacity $x$ to this hypermultiplet, and compute the first few terms of the index of $E_{6, 2}$.   

Identifying $x= e^{\half \b (\e_1 - \e_2)}, \tau = e^{\half \b(\e_1 + \e_2)}$, we find that the superconformal index agrees with our computation. Note that for one instanton, we have to multiply the center of mass contribution to the partition function 
\be
Z_{NY} = \frac{I_{HL}(E_{r, 1})}{(1- e^{\b \e_1})(1 -e^{\b \e_2})} \ ,
\ee
since the index captures only the Hilbert series of the centered instantons.
In the case of $n \ge 2$, the index of \cite{Gaiotto:2012uq} include the center of mass contribution,
\be
Z_{NY} = I_{HL}(E_{r, n})\ .
\ee
Our computation supports the lifting box prescription to obtain the index of non-Lagrangian theories with decoupled hypermultiplets and also confirms that the instanton moduli space consists of the Higgs branch of the whole theory including the decoupled hypermultiplet. We have checked this relation explicitly for $E_{6,2}$ and $E_{7, 2}$ up to the first few orders in $\tau$.   

\section{Conclusion}

In this paper, we have obtained the instanton partition function for $\CN=2$ pure Yang-Mills theory with arbitrary gauge group using the recursion relation of Nakajima-Yoshioka. We re-derived the one-instanton expression of \cite{Keller:2011ek} using the recursion relation, and checked that the 2-instanton partition function agrees with the superconformal index computation proposed in \cite{Gaiotto:2012uq}, which provides a nice consistency check for the index formula of strongly coupled superconformal field theories. We can also write down instanton partition functions of $F_4$ and $G_2$, for which no expressions beyond 1-instanton level were known. Our character expansion \eqref{Z2general} should give hints for the general structure of the instanton moduli space for exceptional groups. 

There are several interesting directions to pursue. First, it is desirable to find an all order expression of \eqref{Z2general} in the vein of (\ref{inst5dk1}), which is universal. It would be nice to have a similar formula for 2-instantons, or more generally for $n$ instantons. This might even make it possible to reverse engineer the moduli space for exceptional gauge groups.

We would also like to mention that our formula can be compared against the norm of certain coherent state of $\CW$-algebra \cite{Alday:2009aq, Gaiotto:2009ma} as in the case of \cite{Keller:2011ek}, so as to provide an additional check of the 2d-4d correspondence. In particular, it should be possible to generalize the relation to the 5d version of the partition function and $q$-deformed $\CW$-algebras \cite{Awata:2009ur, Awata:2010yy}.

\acknowledgments

We would like to thank Abhijit Gadde, Amihay Hanany, Noppadol Mekareeya, and Yuji Tachikawa for discussions and correspondence,
and Davide Gaiotto and Shlomo Razamat for sharing a preliminary draft of \cite{Gaiotto:2012uq} with us. We particularly thank Yuji Tachikawa for introducing to us the blow-up formula of Nakajima-Yoshioka. 
JS thanks the hospitality of Korea Institute for Advanced Study for hospitality.
The work of CAK is supported by a John A.~McCone Postdoctoral Fellowship.
This work is supported in part by the DOE grant DE-FG03-92-ER40701.

\bibliographystyle{ytphys}
\bibliography{ref}

\end{document}